\documentclass[
  reprint,
  superscriptaddress,
  amsmath,amssymb,
  aps
]{revtex4-2}

\usepackage{graphicx}
\usepackage{courier}
\usepackage{amssymb}
\usepackage{mathtools}
\usepackage{amsmath}
\usepackage{xcolor}
\usepackage{float}
\usepackage{todonotes}

\begin{document}

\title{Quantifying and minimizing dissipation in a non-equilibrium phase transition}

\author{Yuejun Shen}
\affiliation{Department of Materials Science and Engineering, Stanford University, Stanford, California 94305, USA}
\affiliation{Stanford Institute for Materials and Energy Sciences, SLAC National Accelerator Laboratory, Menlo Park, California 94025, USA}

\author{Zhiqiao Jiang}
\thanks{These authors contributed equally to this work.}
\affiliation{Department of Chemistry, Stanford University, Stanford, California 94305, USA}
\affiliation{Department of Materials Science and Engineering, Stanford University, Stanford, California 94305, USA}

\author{Yunfan Huang}
\thanks{These authors contributed equally to this work.}
\affiliation{Faculty of Engineering, Universiti Malaya, Kuala Lumpur, 50603, Wilayah Persekutuan, Malaysia}

\author{Brittany M. Cleary}
\affiliation{Department of Chemistry, Stanford University, Stanford, California 94305, USA}

\author{Yixing Jiang}
\affiliation{Department of Biomedical Data Science, Stanford University, Stanford, California 94305, USA}

\author{Grant M. Rotskoff}
\affiliation{Department of Chemistry, Stanford University, Stanford, California 94305, USA}

\author{Aaron M. Lindenberg}
\email{Contact author:aaronl@stanford.edu}
\affiliation{Department of Materials Science and Engineering, Stanford University, Stanford, California 94305, USA}
\affiliation{Stanford Institute for Materials and Energy Sciences, SLAC National Accelerator Laboratory, Menlo Park, California 94025, USA}
\affiliation{Stanford PULSE Institute, SLAC National Accelerator Laboratory, Menlo Park, California 94025, USA}

\date{\today}

\begin{abstract}
    In a finite-time continuous phase transition, topological defects emerge as the system undergoes spontaneous symmetry breaking. The Kibble-Zurek mechanism predicts how the defect density scales with the quench rate. During such processes, dissipation also arises as the system fails to adiabatically follow the control protocol near the critical point. Quantifying and minimizing this dissipation is fundamentally relevant to nonequilibrium thermodynamics and practically important for energy-efficient computing and devices. However, there are no prior experimental measurements of dissipation, or the optimization of control protocols to reduce it in many-body systems. In addition, it is an open question to what extent dissipation is correlated with the formation of defects. Here, we directly measure the dissipation generated during the voltage-driven Fréedericksz transition of a liquid crystal with a sensitivity equivalent to a $\sim10$ nanokelvin temperature rise. We observe Kibble-Zurek scaling of dissipation and its breakdown, both in quantitative agreement with existing theoretical works. We further implement a fully automated \textit{in-situ} optimization approach that discovers more optimal driving protocols, reducing dissipation by a factor of three relative to a simple linear protocol.

\end{abstract}

\maketitle



\textit{Introduction.---}In a continuous phase transition, the relaxation time of the system diverges near the critical point—a phenomenon known as critical slowing down \cite{Book_critical_phenomena}. When a control parameter is varied at a finite rate across the critical point, the system fails to follow adiabatically due to this divergence. Consequently, different regions in the system independently choose among the degenerate ground states of the symmetry-broken phase, resulting in the formation of spatial domains. The mismatch between domains leads to the formation of topological defects. 

Originally proposed in cosmology, the Kibble-Zurek mechanism (KZM) predicts how the density of topological defects scales with the quench rate of the control parameter \cite{JPA_Kibble_cosmic_string_1976, Nature_Zurek_superfluid_helium_1985, KZM_review_Zurek}. 
KZM has been experimentally confirmed in many classical systems, including multiferroics \cite{PRL_multiferroic_KZM_2012} and liquid crystals \cite{Science_LC_Cosmology_1991, CPC_KZM_LC_2017}. In addition, non-equilibrium critical dynamics have been extensively investigated in quantum systems \cite{PRL_defect_nonlinear_quenching_2008,PRL_optimal_control_quantum_KZM_2008,PRL_adiabatic_passage_QCP_avoid_KZM_2012,PRB_opti_control_reducing_pieces_2013,PRA_nonadiabatic_opti_QCP_2025}, where deviations from the conventional KZM scaling have recently been observed \cite{Nature_quantum_coarsening_Lukin_2025,Nature_quantum_simulator_Google_2025}.

From another perspective, finite-rate driving across phase transition inevitably produces dissipation, as the system lags behind the equilibrium state of the control parameter \cite{EPL_Vaikuntanathan_dissipation_and_lag_2009}. More broadly, such dissipation is present in any non-quasi-static process and becomes especially problematic in computation, where speed is critical. For example, enabling fast bit erasure requires paying an additional work cost due to the thermodynamic irreversibility, described by the finite-time Landauer principle \cite{PRL_finite_time_Landauer_2020}. This has motivated theoretical advances in quantifying and minimizing dissipation through optimal control \cite{PRL_universal_bound_bit_reset_2021,JPC_D_Sivak_opti_control_stochastic_thermodynamics_2023,RPP_driving_rapidly_in_control_2023,PRL_opti_finite_time_stochastic_thermodynamics_2007,PRL_Sivak_Crook_2012,PRE_Grant_opti_protocol_ising_flip_2015,PRE_minimal_diss_far_from_equi_Deffner_2018}. In addition, it is also interesting to study the correlation between dissipation and defect production.

Despite its fundamental and practical importance, measurements of dissipation, especially in many-body systems, remain largely unexplored \cite{PRXLife_pattern_irreversibility}. Experimental implementation of optimal control in such systems is even more uncommon. The primary challenge is that dissipation is often extremely small, making it difficult to detect with standard thermometric methods \cite{Nature_thermometry_nanoscale_2016}. In this work, we present a method to directly measure dissipated energy during the voltage-driven Fréedericksz transition in a nematic liquid crystal (LC). The Fréedericksz transition is the field-induced reorientation of the director above a critical threshold, a fundamental mechanism behind liquid crystal display technology \cite{Book_LC_device_2014}. Our approach enables quantitative measurement of dissipation as a function of quench rate, with a sensitivity equivalent to a temperature increase on the order of $10\,$nK in an isolated system.


\begin{figure}[h!]
    \includegraphics[width=\columnwidth]{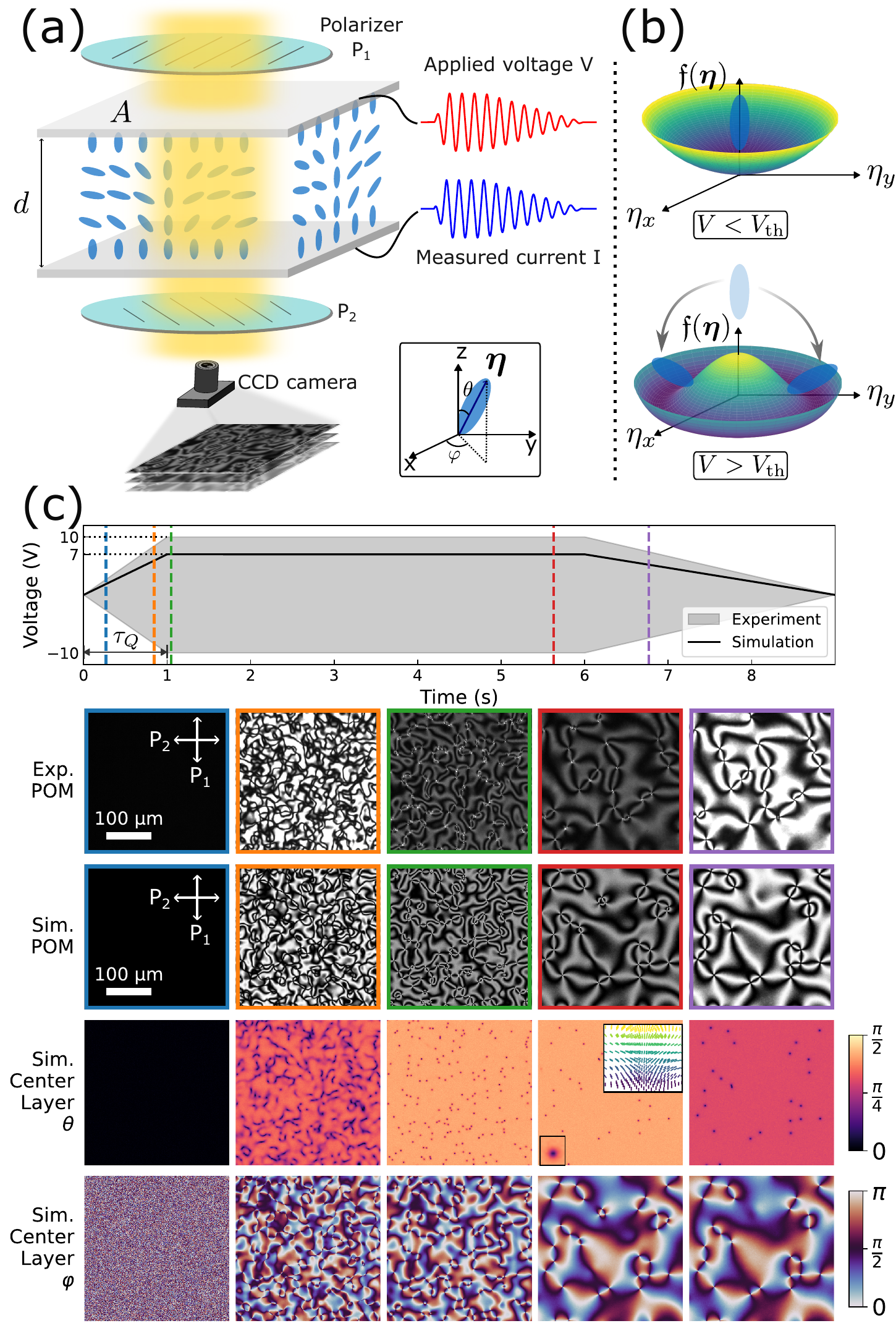}
    \caption{\textbf{Fréedericksz transition of MBBA under an applied voltage.} \textbf{(a)} The LC cell is imaged under a polarized optical microscope (POM) with crossed polarizers $P_1$ and $P_2$. Representative POM images are shown in the image stack. A bipolar voltage ($1$kHz carrier wave) with ramp time $\tau_Q$ is applied to the cell and the current is measured. The LC director is denoted by the vector field $\boldsymbol{\eta}$ with unit norm. \textbf{(b)} Below $V_{\mathrm{th}}$, the vertically aligned configuration minimizes the free energy. Above $V_{\mathrm{th}}$, the director tilts toward the $x-y$ plane and locally chooses a specific azimuthal angle. \textbf{(c)} Colored dotted lines on the voltage trace indicate the acquisition times of the images, with colors matching the borders of the POM frames. The insets in the fourth panel of the third row show magnified views of a topological defect in 2D (view along $z$) and 3D (view along $x$). }
    \label{Fig1}
\end{figure}


\textit{Experiment and simulation setup.---} As illustrated in Fig.~\ref{Fig1}(a), an Instec homeotropic liquid crystal (LC) cell ($d = 5\,\mu\text{m}$, $A = 35\,\text{mm}^2$) is filled with N-(4-methoxybenzylidene)-4-butylaniline (MBBA) with negative dielectric anisotropy \cite{Book_Phy_LC}. A bipolar voltage signal with ramp time $\tau_Q$ is applied to the cell and the current is measured. The cell is imaged under a polarized optical microscope (POM). The local molecular orientation is described by the director $\boldsymbol{\eta}(x,y,z)$. The experimental details are in  Supplementary Note 1.

We define the system to be the LC layer under a uniform electric field $\boldsymbol{E}=E\hat{z}=(V/d)\hat{z}$, which acts as the control parameter. The free energy density of the system at position $\boldsymbol{r}$ is given by \cite{Book_Phy_LC}:
\begin{align} \label{f}
    \mathfrak{f(\boldsymbol{r},\boldsymbol{\eta})}= \frac{\mathcal{K}}{2} \left|\nabla \boldsymbol{\eta}(\boldsymbol{r})\right|^2 -\frac{1}{2}\varepsilon_0\Delta \varepsilon \left(\boldsymbol{E} \cdot \boldsymbol{\eta}(\boldsymbol{r}) \right)^2,
\end{align}
where $\mathcal{K}$ is the Frank elastic constant with one-constant approximation and $\Delta \varepsilon \coloneq \varepsilon_{\parallel}- \varepsilon_{\perp}<0$ is the dielectric anisotropy of MBBA.

When $V=0$, the director aligns uniformly along the $z$ axis due to the homeotropic boundary conditions. As $V$ slowly increases, the director remains undistorted until the threshold $V_{\mathrm{th}}$, where the decrease in electrical free energy outweighs the elastic penalty. $V_{\mathrm{th}}=6.1$V is optically measured using a very slow voltage ramp (Fig. S1). As shown in Fig. \ref{Fig1}(b), the stable configuration switches from vertical alignment ($\theta=0$) to a tilt state ($\theta>0$) as $V>V_{\mathrm{th}}$. The degeneracy of the azimuthal angle $\varphi$ leads to a ``Mexican hat" energy landscape. As a result, the director locally chooses a specific $\varphi$, spontaneously breaking the system's rotational symmetry about the $z$ axis, and leading to the formation of topological defects according to KZM. 

To model the Fréedericksz transition, we use the model A dynamics with a stochastic thermal noise \cite{Bray_phase_ordering_kinetics}. The details of the simulation are in the Supplementary Note 2. Fig. \ref{Fig1}(c) compares experimental and simulated behavior of the LC under an applied voltage. In the simulation, a DC voltage protocol is used, with the applied voltage set to the RMS value of the experimental waveform. At low voltage, the director remains vertically aligned, resulting in a dark POM image. The orange dotted line marks the time when the director begins tilting toward the $x-y$ plane. The voltage at this point ($\approx 8$V) is higher than the quasi-static threshold $V_{\mathrm{th}}=6.1$V as the system lags behind the fast ramping voltage. Topological defects become well-resolved around $t=1$s, followed by defect annihilation. As the voltage ramps down, the director returns toward vertical alignment. The simulated POM images closely match experimental observations at all time points, validating the model. The third and fourth row show the simulated polar $\theta$ and azimuthal $\varphi$ angles of the center layer director, corresponding to the simulated POM image above. Topological defects appear at the intersections of domains, marked by a vertical director at the defect core. The $\pm 2 \pi$ winding in $\varphi$ indicates a topological charge of  $\pm 1$. A 3D visualization of the director configuration near a defect is shown in the inset. The full video of Fig. \ref{Fig1}(c) is in Supplementary Video 1.

\begin{figure}[h!]
    \centering
    \includegraphics[width=\columnwidth]{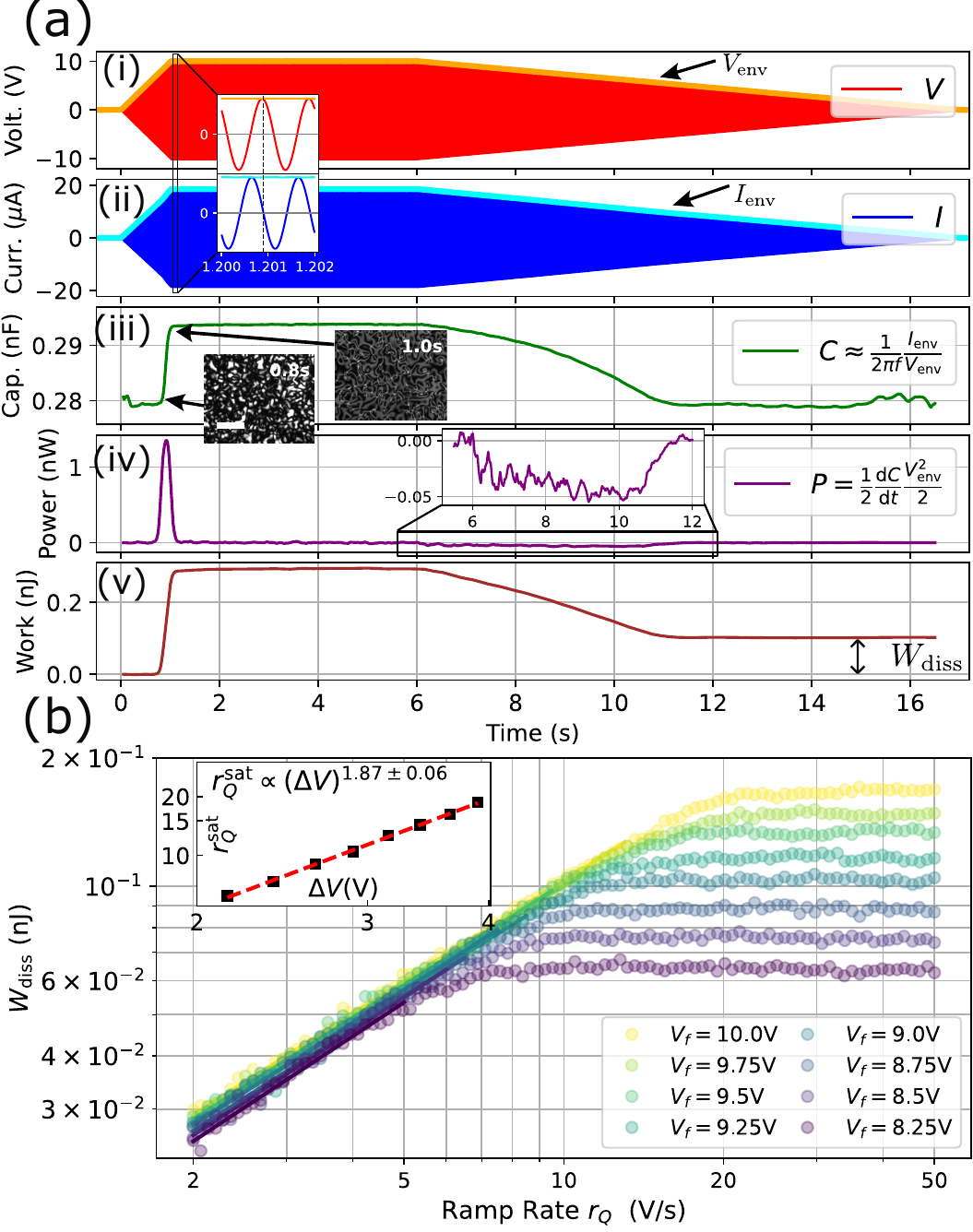}
    \caption{\textbf{Experimental measurement of dissipated work.} \textbf{(a)} $V_{\mathrm{env}}$ and $I_{\mathrm{env}}$ are the envelope amplitudes of voltage $V$ and current $I$. The inset magnifies the boxed region, showing a phase lag of the current behind the voltage close to $\pi/2$. The green curve shows the calculated capacitance of the LC cell. The scale bar of the POM images is 100\,$\mu$m. The purple curve shows the relevant input power. The inset highlights the time of negative power input.  The brown curve shows the relevant work done $W$. \textbf{(b)} Linear regions in the log-log plot are fitted with power laws (shown in lines), obtaining similar coefficient $\kappa$ between 0.79 and 0.82. The inset (top left) plots the saturation ramp rate $r_Q^{\mathrm{sat}}$ against $\Delta V = V_f - V_{\mathrm{th}}$ in log-log scale. The fitted power law is shown in text.}
    \label{Fig2}
\end{figure}

\textit{Kibble-Zurek scaling and its breakdown.---} We show in Supplementary Note 3 that the Fréedericksz transition for the case $\Delta \varepsilon<0$ is a second-order phase transition, similar to those with $\Delta \varepsilon>0$ \cite{PRL_nonGaussian_fluctuation_LC_freedericks,Book_Phy_LC}. The in-plane correlation length $\xi$ and susceptibility $\chi$ near the threshold scale as: 
\begin{align}
    \xi \sim \left(\Delta E\right)^{-\nu}&, \nu=\frac{1}{2}; \\
    \chi \sim \left(\Delta E\right)^{-\Lambda}&, \Lambda=\frac{1}{2}.
\end{align}
Here $\Delta E \coloneq E-E_{\mathrm{th}}$, where $E_{\mathrm{th}} \coloneq V_{\mathrm{th}}/d$. The dynamic critical exponent $z=2$ for model A dynamics \cite{RMP_Hohenberg_Halperin_1977, PhysicaA_ModelA_z_2022}. Therefore, according to KZM, the defect number $N$ scales with the ramp time $\tau_Q$ as \cite{KZM_review_Zurek}:
\begin{align}
    N \sim \left(\frac{1}{\tau_Q}\right)^\mu,\mu={\frac{(D_{\mathrm{sys}}-D_{\mathrm{def}})\nu}{1+z\nu}}=\frac{1}{2}
\end{align}
Here, $D_{\mathrm{sys}}=3$ is the system dimension and $D_{\mathrm{def}}=1$ is the defect dimension. 

In the LC cell, the thick glass substrates surrounding the LC layer serve as a thermal reservoir at $T= 23 ^\circ \mathrm{C}$. In such an isothermal process, the dissipated energy is equal to the dissipated work, defined as the difference between the total work done to the system $W$ and the free energy change of the system $\Delta \mathcal{F}$ \cite{Book_callen2006thermodynamics}:
\begin{align}\label{W_diss_def}
    W_{\mathrm{diss}}\coloneq W-\Delta\mathcal{F}
\end{align}
$W_{\mathrm{diss}}$ is predicted to scale with the ramp time $\tau_Q$ as \cite{PRE_KZM_entropy}:
\begin{align}
    W_{\mathrm{diss}} \sim \left(\frac{1}{\tau_Q}\right)^\kappa, \kappa={\frac{2-\Lambda}{1+z\nu}}=\frac{3}{4}
\end{align}

Our main goal is to experimentally extract the dissipated work $W_{\mathrm{diss}}$ as the voltage ramps from 0 to $V_f$ ($V_f>V_{\mathrm{th}}$) over a finite ramp time $\tau_Q$. Direct measurement of $W$ is experimentally accessible, whereas determining $\Delta \mathcal{F}$ is considerably more challenging. To eliminate $\Delta \mathcal{F}$ in Eq. \eqref{W_diss_def}, we implement a voltage cycle that starts and ends at zero voltage, thereby returning the system to its original equilibrium state and ensuring $\Delta \mathcal{F} = 0$. In this case, the dissipated work is equal to the total work done over the cycle.

An example voltage control protocol $V(t)$ with $\tau_Q=1$s is shown in the first panel of Fig. \ref{Fig2}(a). The voltage is linearly increased from $0$ to $V_f=10$V over $0<t\leq 1$s, held constant at $V_f=10$V for $\tau_{V_f}=5$s ($1\mathrm{s}<t\leq 6\mathrm{s}$) and then linearly decreased back to 0 over $\tau_{\mathrm{down}}=11$s ($6\mathrm{s}<t\leq 17\mathrm{s}$). The plateau duration $\tau_{V_f}$ allows the system to mostly relax toward the equilibrium configuration at $V_f$, such that the majority of dissipation associated with relaxation has already occurred. The slow ramp-down ($\tau_{\mathrm{down}} \gg \tau_Q$) ensures that the dissipation during the downward transition is negligible compared to that generated during the ramp-up transition. As a result, the dissipated work measured over the full cycle is, to a good approximation, equal to the dissipation generated during the ramp-up process $\tau_Q$. Figure. S2 supports this by showing that the measured dissipation has already converged as $\tau_{V_f}$ and $\tau_{\mathrm{down}}$ exceed 5s and 11s, respectively. These durations in the protocol therefore offer a practical balance between measurement accuracy and total cycle time.

To extract the total work done $W$, we measure the real-time current through the cell $I$ (Fig.~\ref{Fig2}(a)(ii)). $V_{\mathrm{env}}$ and $I_{\mathrm{env}}$ are the envelope of the $V$ and $I$, respectively. Directly calculating work done as $W=\int dt VI$ is not appropriate in this context, as the result would be dominated by resistive heating due to impurity ion motion. To isolate the dissipation associated with LC molecular rotation, we first estimate the cell capacitance $C \approx I_{\mathrm{env}}/(2\pi f V_{\mathrm{env}})$ (Fig. \ref{Fig2}(a)(iii)). By only considering the LC rotational component that contributes to dissipation under this protocol, we calculate the relevant power as $P=\frac{1}{2} \frac{\mathrm{d}C}{\mathrm{d}t}\frac{V^2_{\mathrm{env}}}{2}$ (Fig. \ref{Fig2}(a)(iv)). The total work done $W$ is given by integrating the power over time (Fig. \ref{Fig2}(a)(v)). The dissipated work $W_\mathrm{diss}$ is equal to the final value of $W$. Details regarding the estimation error of $C$ and the derivation of the relevant power expression are in Supplementary Note 4.


Using the methods described above to extract $W_{\mathrm{diss}}$, we now investigate its Kibble-Zurek scaling. For a linear ramping protocol, the ramp rate is defined as $r_Q \coloneq V_f / \tau_Q$. As shown in Fig. \ref{Fig2}(b), under fixed $V_f$,  $W_{\mathrm{diss}}$ exhibits a power-law dependence on $r_Q$ in the low $r_Q$ regime. The data indicates that $W_{\mathrm{diss}}$ can be resolved to better than $0.01\,$nJ, corresponding to less than $30\,$nK temperature change in the LC, if thermally isolated (details in Supplementary Note 5). The experimental scaling exponent $\kappa \approx 0.8$ is in close agreement with the prediction $\kappa=0.75$ ($\kappa$ for each $V_f$ curve is available in Table S1). At higher $r_Q$, $W_{\mathrm{diss}}$ saturates beyond a characteristic value $r_Q^{\mathrm{sat}}$, signaling the breakdown of KZM scaling as predicted theoretically in previous work \cite{PRL_breakdown_KZM_2023}. $r_Q^{\mathrm{sat}}$ is defined as the ramp rate at which the extrapolated low-$r_Q$ power-law fit reaches the saturation value of $W_{\mathrm{diss}}$. When varying $V_f$ (while keeping $V_f > V_{\mathrm{th}}$), two notable features emerge. First, all curves coincide in the low-$r_Q$ regime. This implies that $W_{\mathrm{diss}}$ is mainly generated when the voltage is near $V_{\mathrm{th}}$, where the relaxation time diverges, and is therefore independent of $V_f$. Second, $r^{\mathrm{sat}}_Q$ is higher for larger $V_f$. The inset of Fig.~\ref{Fig2}(b) shows $r_Q^{\mathrm{sat}}$ as a function of $\Delta V \coloneq V_f - V_{\mathrm{th}}$, consistent with the theoretical prediction $r_Q^{\mathrm{sat}} \sim \Delta V^{1+z\nu}$ \cite{PRL_breakdown_KZM_2023}, with $1 + z\nu = 2$ for this system. The universality of the power law exponent $\kappa$ is further verified at an elevated temperature $29\, ^\circ \mathrm{C}$ (Fig. S5).


To study the defect number $N$, we designed an unbiased counting method based on POM image analysis. We observe a Poisson spatial distribution of topological defects and reveal a slight deviation from the KZM prediction on the scaling of $N$ versus $r_Q$, likely due to the defect annihalation dynamics. Detailed analyses are in Supplementary Note 6.

Fig. \ref{Fig3}(a) shows the synthetic measurements from simulation over a $100\,\mu \mathrm{m} \times100\,\mu \mathrm{m}$ area (details in Supplementary Note 2). The dynamics of $C_{\mathrm{meas}}$, $P_{\mathrm{meas}}$, and $W_{\mathrm{meas}}$ show excellent quantitative agreement with the experimental results in Fig.~\ref{Fig2}(a), after applying the $3500\times$ area scaling factor. In addition, the true dissipation dynamics  $W_{\mathrm{diss,true}}(t')=-\int_{t_0}^{t'}\mathrm{d}t \ \frac{\delta \mathcal{F}}{\delta \boldsymbol{\eta}} \frac{\partial \boldsymbol{\eta}}{\partial t}$ \cite{Bray_phase_ordering_kinetics} are presented together with $W_{\mathrm{meas}}$ in Fig. \ref{Fig3}(a)(iv). Here, $\mathcal{F}$ is the total free energy defined as $\mathcal{F}\coloneq\int \mathrm{d}\boldsymbol{r} \mathfrak{f}(\boldsymbol{r},\boldsymbol{\eta})$. $W_{\mathrm{diss,true}}$ increases monotonically as expected from the second law of thermodynamics, with the steepest growth occurring at the phase transition during the ramp-up stage. At the end of the measurement, $W_{\mathrm{diss,true}}$ and $W_{\mathrm{meas}}$ closely agree, confirming that the experimental method accurately extracts the true dissipation.  Fig. \ref{Fig3}(b) shows the Kibble-Zurek scaling and its breakdown. The extracted power-law exponents agree closely with those in Fig. \ref{Fig2}(c), further supporting the model's validity in describing the system (fitted $\kappa$ for each $V_f$ is in Table S1).

\begin{figure}[h]
    \includegraphics[width=\columnwidth]{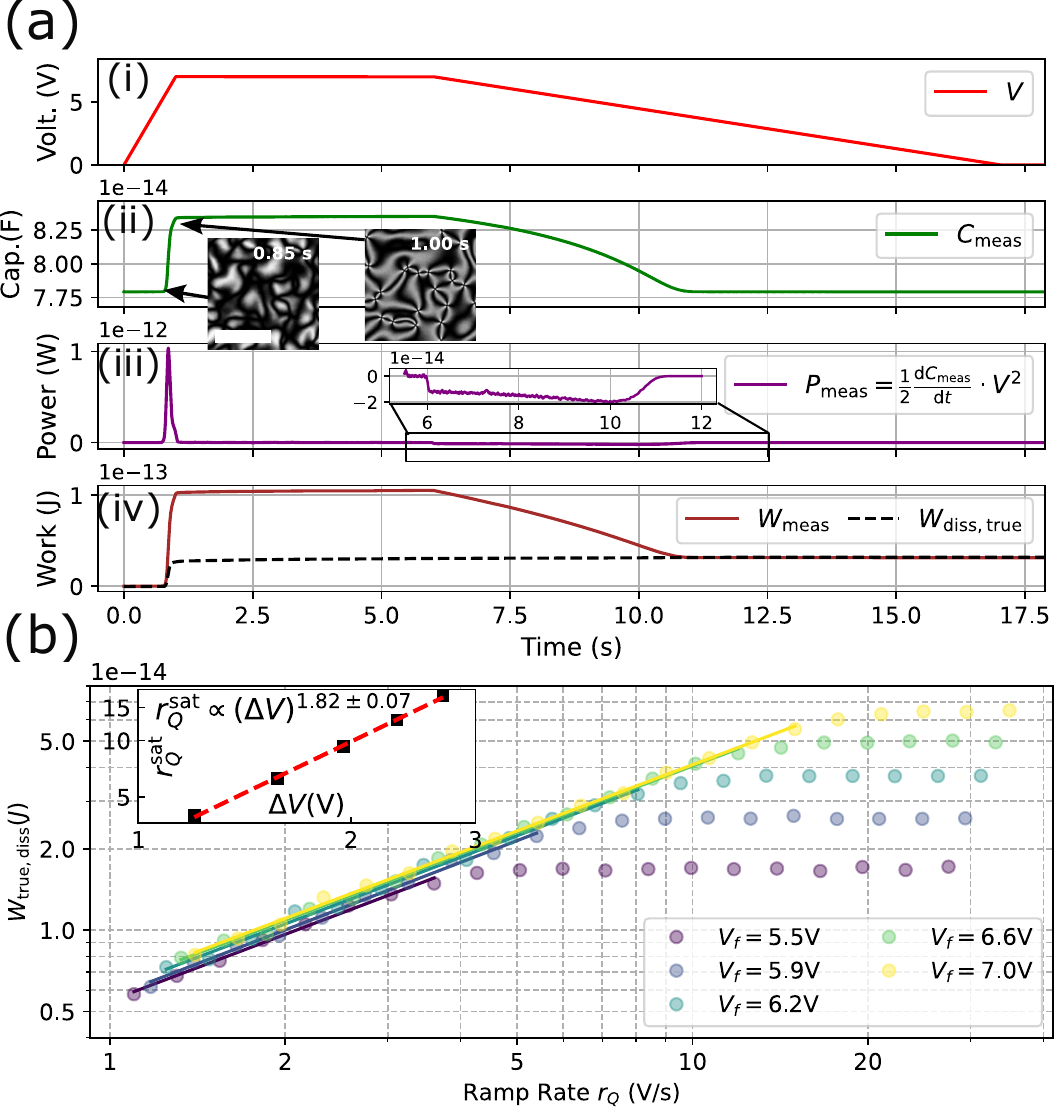}
    \caption{\textbf{Simulated dissipated work as the voltage ramps through the threshold.} \textbf{(a)} Simulation counterpart to the experimental data in Fig. \ref{Fig2}(a). The scale bar of the POM image represents 50$\mu$m. \textbf{(b)} Simulation counterpart to the experimental data in Fig. \ref{Fig2}(b). The power law coefficient $\kappa$ is between 0.81 and 0.82 for different $V_f$.}
    \label{Fig3}
\end{figure}

\begin{figure}[h!]
    \includegraphics[width=1\columnwidth]{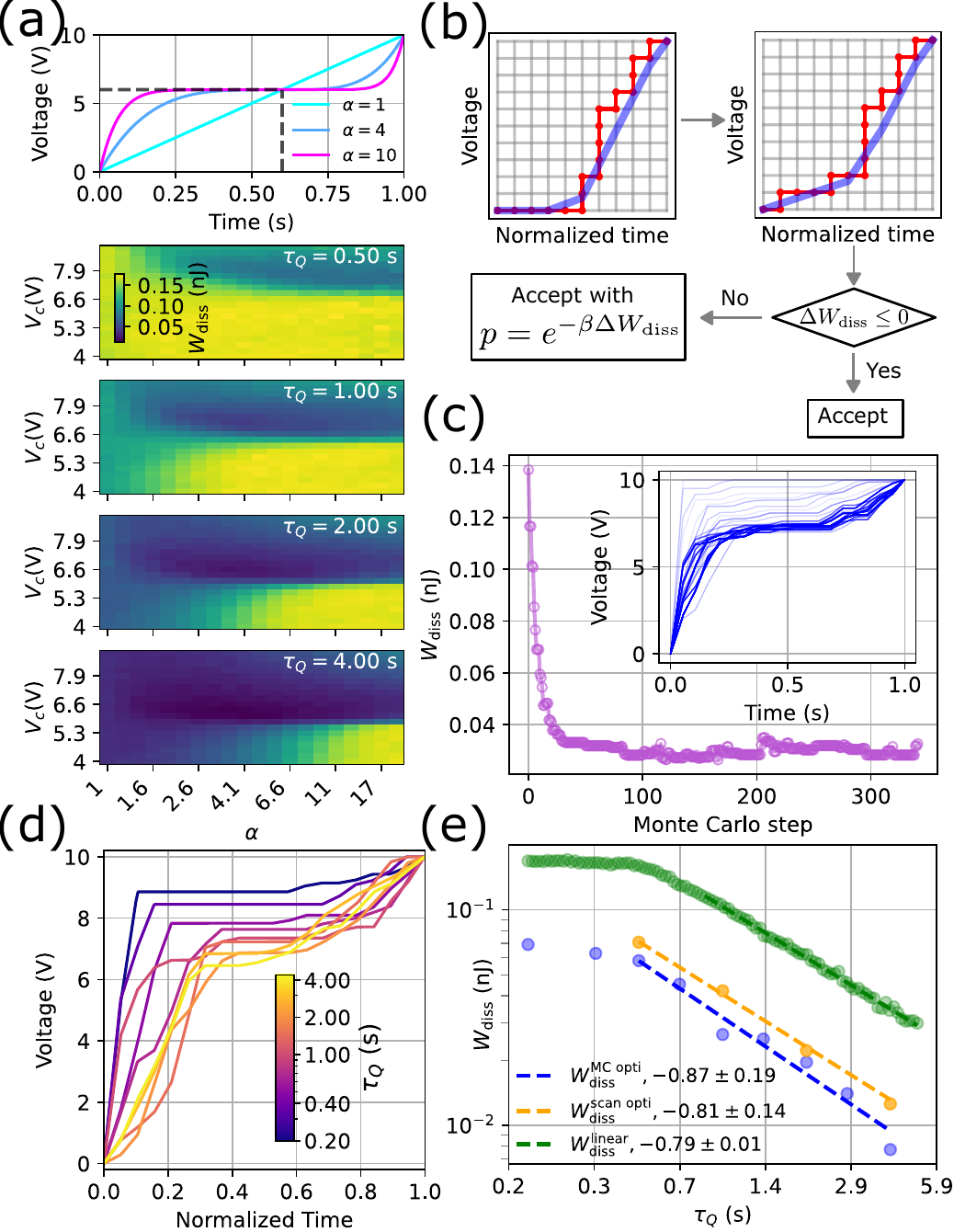}
    \caption{\textbf{Experimental optimization of dissipated work.} \textbf{(a)} Top panel: $\tau_Q=1$s and $V_c=6$V, varying $\alpha$. Colormaps (share colorbar): $W_{\mathrm{diss}}$ as a function of $(\alpha, V_c)$ for different $\tau_Q$. All voltages here and below refer to $V_{\mathrm{env}}$ of the protocol. \textbf{(b)} MCMC workflow for protocol optimization. The grid shown in schematics is for illustration (actual size $40\times40$). Red: discretized protocol. Blue: smoothed protocol. The pink arrow shows a random swap. \textbf{(c)} Example MCMC optimization trajectory for $\tau_Q = 1$s and $V_f=10$V. The inset shows the protocol trajectories in faint blue. The darker the curve, the more frequently the protocol appears. \textbf{(d)} Optimal protocols for various $\tau_Q$, plotted with the time axis normalized by $\tau_Q$. \textbf{(e)} Dissipated work under different optimizations. Green: linear ramp. Orange: best protocol from parameter scan in (a). Blue: MCMC-optimized protocol from (d). Fitted power-law exponents are shown in the legend.}
    \label{Fig4}
\end{figure}

\textit{Optimization.---} In this section, we seek an optimal voltage protocol whose envelope ramps from $0$ to $V_f$ in time $\tau_Q$ while minimizing dissipation. Since most dissipation occurs near the critical point, a linear ramp is unlikely to be optimal. Intuitively, the optimal protocol should adjust its rate of change to follow the system's relaxation rate, slowing down as it approaches the critical point and accelerating when far from it. Therefore, we first assume the optimal protocol to be parameterized as: 
\begin{align}
V(t) =
\begin{cases}
V_c + A_1 |t - t_c|^{\alpha}, & t \leq t_c, \\
V_c + A_2 |t - t_c|^{\alpha}, & t > t_c .
\end{cases}
\end{align}
where $V_c$ is the critical voltage and $t_c=\tau_Q V_c/V_f$ is the time at which $V_c$ is reached. The constant $A_1$ and $A_2$ are chosen to satisfy the constraints $V(0)=0$ and $V(\tau_Q)=V_f$. The exponent $\alpha$ controls the flatness of the ramp near $V_c$. 

The first panel of Fig. \ref{Fig4}(a) shows examples of parameterized control protocol. We perform a two-dimensional scan over $(\alpha, V_c)$ for several ramp times $\tau_Q$, with the results shown as the four colormaps in Fig. \ref{Fig4}(a). These results confirm our earlier expectation that the linear protocol ($\alpha = 1$) is not optimal. The optimal parameters $\alpha^{\mathrm{opt}}$ and $V_c^{\mathrm{opt}}$ both increase as $\tau_Q$ decreases. A similar scan to minimize the number of defects (Fig.~S12) shows a distinct difference, including much smaller changes in $V_c^{\mathrm{opt}}$ with respect to $\tau_Q$ and a much broader range of $\alpha^{\mathrm{opt}}$ for long $\tau_Q$. Despite complications arising from defect annihalation, this result qualitatively indicates a complex correlation between defect formation and dissipation \cite{PRE_KZM_entropy}. In addition, we note that our results differ from the theoretical prediction of Ref. \cite{PRL_optimal_control_quantum_KZM_2008}, which suggests that $\alpha_{\mathrm{opt}}$ should increase while $V_{\mathrm{opt}}$ remains constant as $\tau_Q$ increases. This discrepancy likely arises because our strong driving violates the adiabatic assumption of Ref. \cite{PRL_optimal_control_quantum_KZM_2008}.

However, parameterizing the protocol by $(\alpha,V_c)$ only represents a limited subset of the protocols that satisfy the constraints. To efficiently search in a broader protocol space, we perform a Markov chain Monte-Carlo (MCMC) search (Fig.~\ref{Fig4}(b)). Any monotonic protocol satisfying the constraints can be discretized as a path on a $m \times m$ grid consisting of $m$ right and $m$ upwards steps, given that $m$ is large enough. Starting from an arbitrary path, the corresponding protocol is obtained via smoothing and its $W_{\mathrm{diss}}$ is measured experimentally. Next, a new candidate protocol is generated by swapping a randomly chosen right step with a randomly chosen upward step. Its $W_{\mathrm{diss}}$ is then measured. If the change $\Delta W_{\mathrm{diss}} \leq 0$, the new protocol is accepted. Otherwise, it is accepted with probability $p=e^{-\beta \Delta W_{\mathrm{diss}}}$. The inverse ``temperature" $\beta$ is tuned to balance convergence and exploration.

Fig. \ref{Fig4}(c) shows an example trajectory of the MCMC optimization. As seen from the inset, a large set of near-optimal protocols is found to yield comparably low $W_{\mathrm{diss}}$, consistent with theoretical predictions in Ref. \cite{PNAS_near_optimal_protocols_spin_flip_2016}. The overall optimal protocol shape qualitatively follows our expectation of having a slower ramp rate near the phase transition. Fig. \ref{Fig4}(d) presents the optimal protocol for different $\tau_Q$. Their qualitative features agree with the parameter scans in Fig.~\ref{Fig4}(a): both the plateau voltage and its flatness increase as $\tau_Q$ decreases. For long $\tau_Q$, the plateau voltage approaches the equilibrium threshold voltage $V_{\mathrm{th}}=6.1$V. In contrast to the assumption of the parameterization, the inflection points of the MCMC-optimized protocols do not lie on the straight line connecting the initial and final   points. Fig.~\ref{Fig4}(e) compares dissipation from the linear protocol, parameterized optimization, and MCMC optimization, showing a progressive decrease across the three cases. The MCMC-optimized protocol reduces the dissipation by a factor of three compared to the linear protocol. At long $\tau_Q$, $W_{\mathrm{diss}}^{\mathrm{MC\ opti}}$ and $W_{\mathrm{diss}}^{\mathrm{scan\ opti}}$ share a similar Kibble-Zurek scaling as the linear protocol.


\textit{Conclusion.---}In this work, we experimentally quantified and minimized the dissipation generated during the non-equilibrium Fréedericksz transition. Dissipation measurement showed $\sim10\,$nK sensitivity and was in close agreement with the predicted Kibble-Zurek scaling as well as its breakdown. We implemented an automated \textit{in-situ} MCMC search to optimize the control protocol. This method produced a more optimal protocol than a parameterized scan, reducing the dissipation by a factor of three compared to the linear protocol. Our results confirm that dissipation and defect production follow different scaling laws and show further that they have different optimal protocols. This work establishes the Fréedericksz transition of LC as a clean, well-controlled platform for studying dissipation in non-equilibrium phase transitions. The spontaneous breaking of continuous symmetry, characterized by a Mexican hat free energy landscape, also offers a useful analogy to study other non-equilibrium phenomena such as the Higgs modes in condensed matter \cite{PRL_Higgs_mode_BCS_Superconductors_Terahertz_2013,NatComm_Higgs_mode_iron_superconductors_2021,Arxiv_Higgs_mode_multiferroic_2025} and cosmic string formation in cosmology \cite{RepProgPhy_Cosmic_String_1995}. This study may also enable new characterization approaches in electronic nematic phases \cite{AnualReview_electron_Nematic_2010}. Finally, our method for extracting dissipation and designing optimal protocols is also applicable for other electrically driven systems such as ferroelectric switching.

\textit{Acknowledgements.---} This work was supported by the US Department of Energy (DOE), Oﬃce of Basic Energy Sciences, Division of Materials Sciences and Engineering, under contract no. DE-AC02-76SF00515.

\bibliographystyle{naturemag}
\bibliography{mybib}

\end{document}